\documentclass[a4paper,12pt,twocolumn]{article}
\usepackage[T1]{fontenc}
\usepackage{comment}
\usepackage{amsfonts}
\usepackage{ae,aecompl}
\usepackage{amsmath}
\usepackage{amssymb}
\usepackage{amsthm}
\usepackage[pdftex]{graphicx}
\usepackage{pstricks}
\usepackage{tikz}
\usepackage{epstopdf}
\usepackage[small]{caption}   
\usepackage{color}
\usepackage{enumerate} 
\usepackage{floatrow}
\usepackage{hyperref}
\usepackage{mathdots}
\usepackage{hyperref}
\hypersetup{
    colorlinks=true,
    linkcolor=blue,
    filecolor=magenta,      
    urlcolor=blue,
}

\definecolor{lightgray}{gray}{0.8}

\title{CT Data of a Pen-Spring: Application to Under-Sampled Dynamic X-ray Tomography}
\author{J. Juurakko,\footnote{Department of Automation and Electrical Engineering, Aalto University, Finland (juliaana.juurakko@aalto.fi)}
\ Z. Purisha,\footnote{Department of Automation and Electrical Engineering, Aalto University, Finland (zenith.purisha@aalto.fi)}  
\ and S. S\"arkk\"a\footnote{Department of Automation and Electrical Engineering, Aalto University, Finland (simo.sarkka@aalto.fi)}}

\date{}
\begin{document}

\maketitle
\abstract{This is the documentation of Computed Tomography (CT) data of a pen-spring. The open data set is available \href{https://zenodo.org/record/3266936#.XRyMdCZS9oA}{here} 
and can be freely used for scientific purposes with appropriate references to the data and to this document in \url{http://arxiv.org/}. The provided data set includes the X-ray sinograms ({\tt finalSino}) of a single 2D slice from a different height of the spring. The {\tt finalSino} was obtained from a measured 10-projection or 100-projection {\tt sinogram} using fan-beam geometry
by down-sampling and taking logarithms. The data set includes also those original measured {\tt sinogram}s and corresponding measurement matrices.}

\section{Introduction}

\begin{figure*}[ht]
\begin{picture}(300,400)
\put(-90,0){\includegraphics[width=17cm]{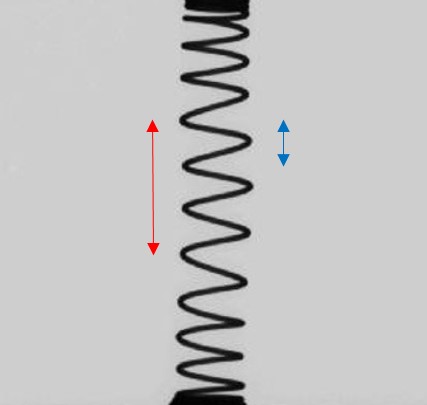}}
\centering
\caption{The blue arrow shows where 11 measurements with 100 projections were taken and the red arrow shows where the 33 measurements were taken with 10 projections.}
\label{fig:springarrows}
\end{picture}
\end{figure*}

\begin{figure*}
\begin{picture}(390,500)
\put(-27,440){\includegraphics[width=150pt]{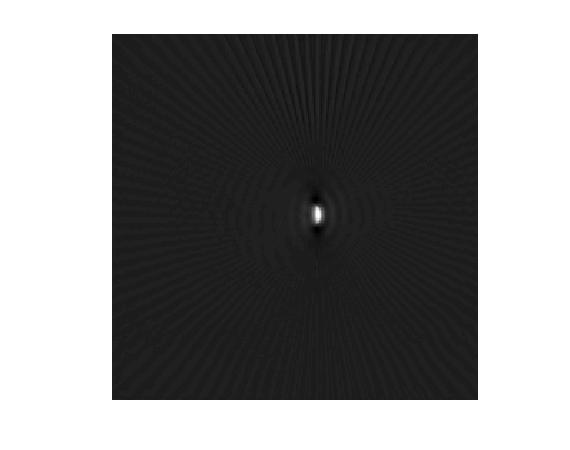}}
\put(123,440){\includegraphics[width=150pt]{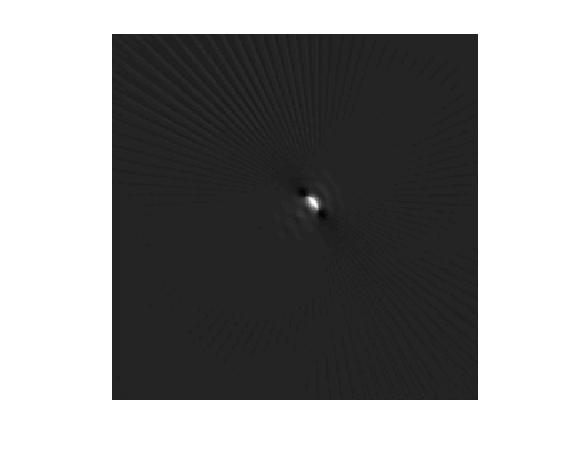}}
\put(273,440){\includegraphics[width=150pt]{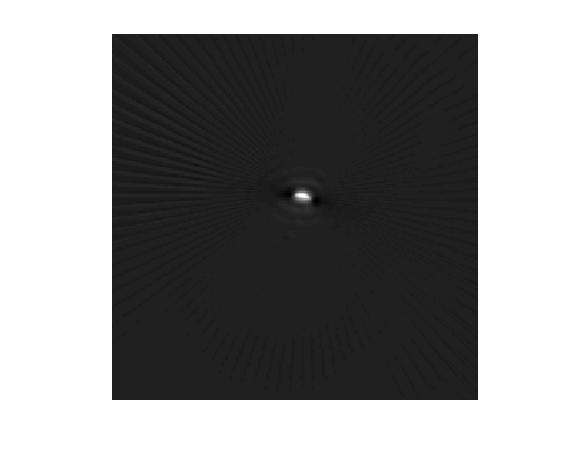}}
\put(-27,300){\includegraphics[width=150pt]{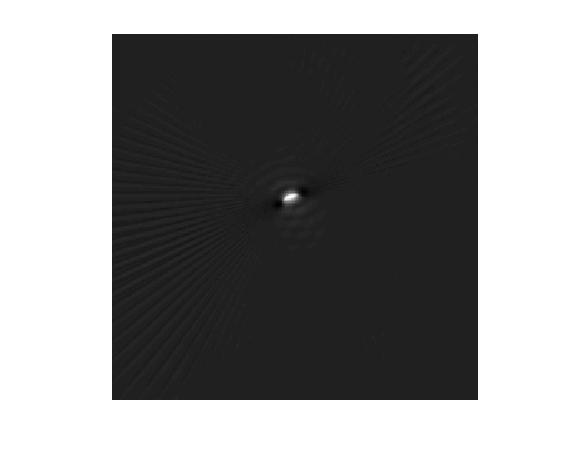}}
\put(123,300){\includegraphics[width=150pt]{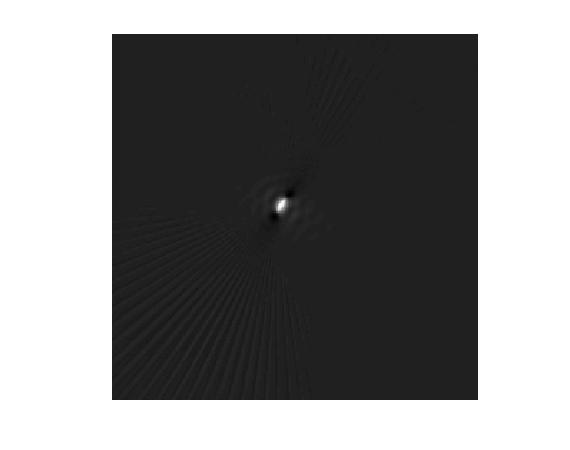}}
\put(273,300){\includegraphics[width=150pt]{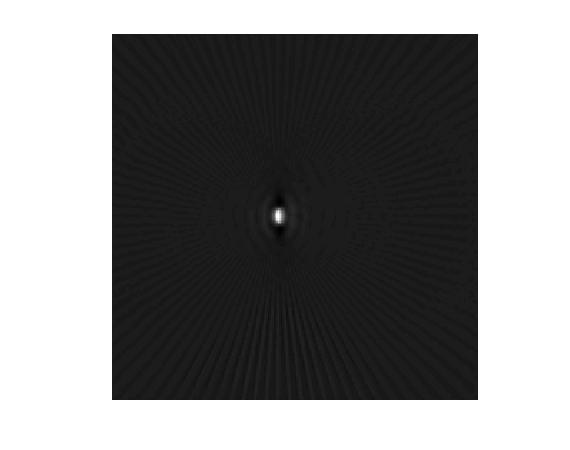}}
\put(-27,160){\includegraphics[width=150pt]{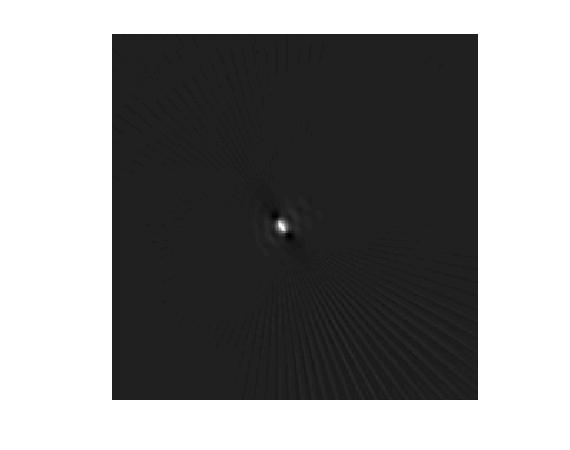}}
\put(123,160){\includegraphics[width=150pt]{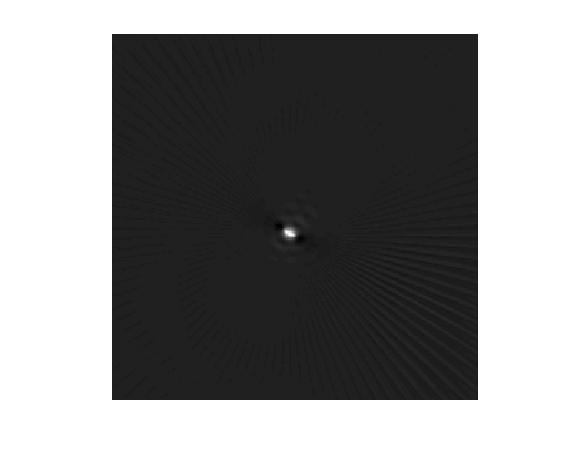}}
\put(273,160){\includegraphics[width=150pt]{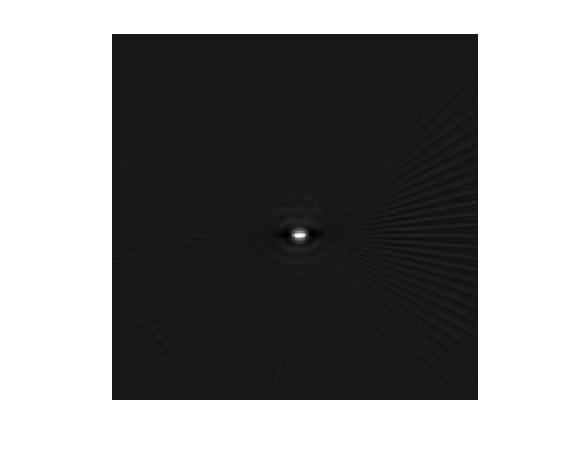}}
\put(-27,20){\includegraphics[width=150pt]{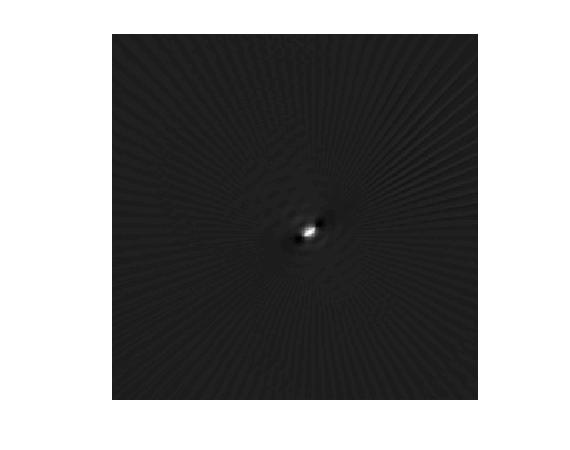}}
\put(123,20){\includegraphics[width=150pt]{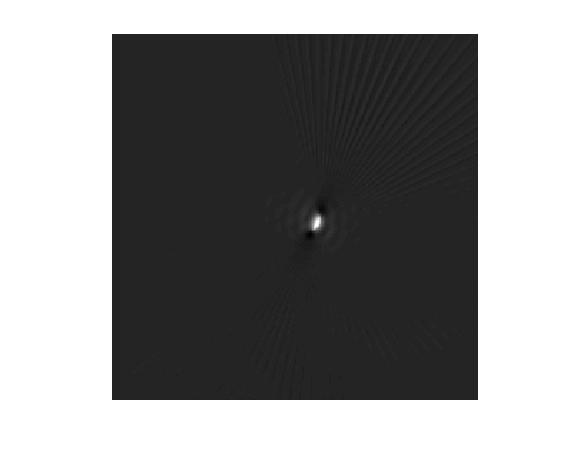}}
\put(42,430){(1)}
\put(192,430){(2)}
\put(342,430){(3)}
\put(42,290){(4)}
\put(192,290){(5)}
\put(342,290){(6)}
\put(42,150){(7)}
\put(192,150){(8)}
\put(342,150){(9)}
\put(42,10){(10)}
\put(192,10){(11)}
\end{picture}
\caption{2D slice reconstructions from 11 different height of the spring, with 100 projections (from {\tt Data\_256x100}).}\label{fig:slices}
\end{figure*}
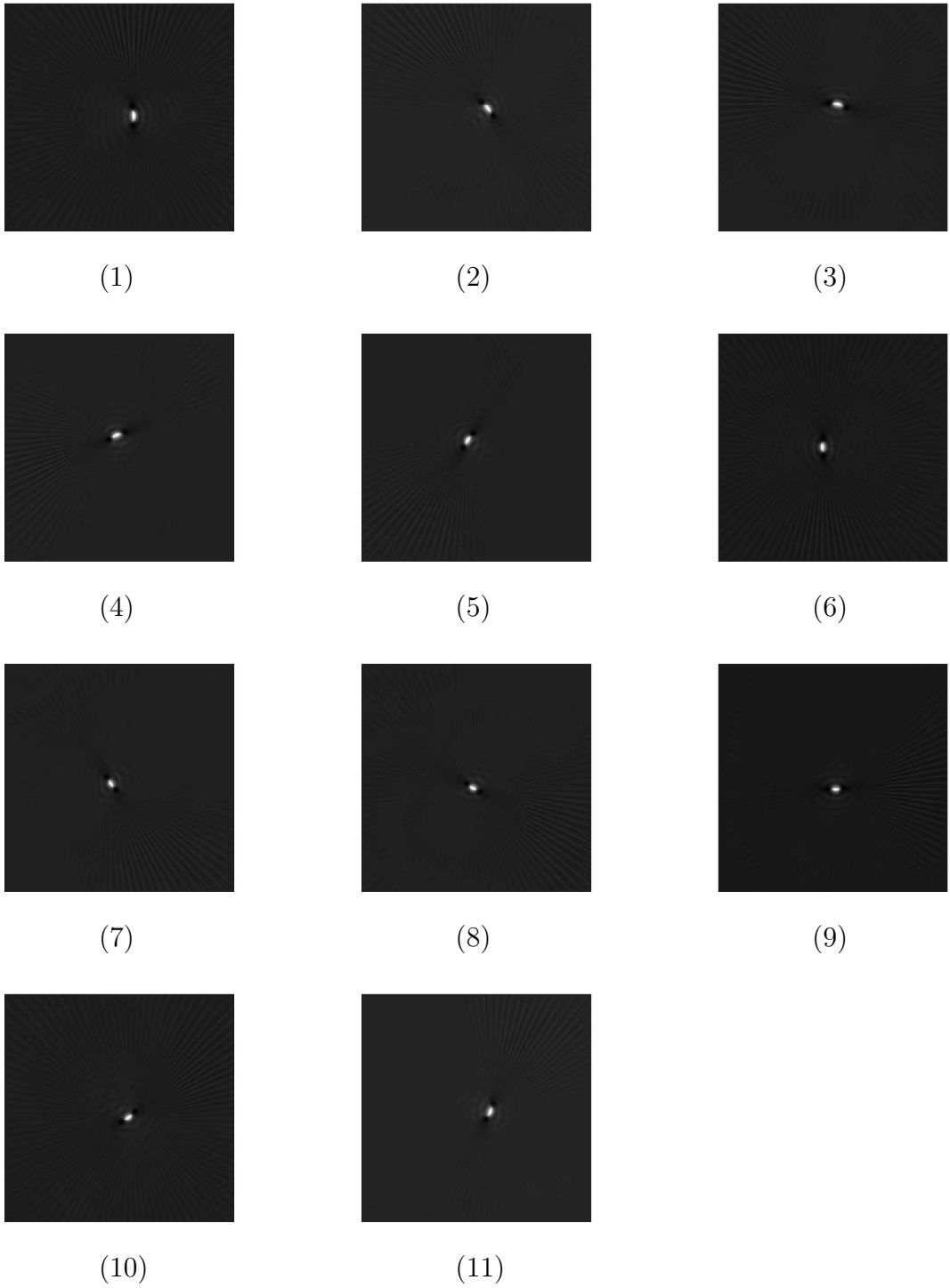

In this documentation, Computed Tomography (CT) data of a pen-spring was acquired. The goal of this data set is to test dynamic x-ray algorithm using under-sampled data. The data is prepared for dynamic case, where the under-sampled data sets are challenging, but also interesting \cite{hakkarainen2019undersampled,bubba2017shearlet,chen2008prior,niemi2015dynamic}. In practice, there are also applications in cardiac CT \cite{hu2012sparse,naoum2015iterative}. Twenty five millimeters long pen-spring with $4$ mm diameter, made from $0,5$ mm thick thread was stretched a little to make the shape less frequent. The object was lifted $0.25$ mm between the measurements to get the full circle from the mid slices of the spring. With 10 projection the spring was measured with 33 different height to get data from three consecutive circles. With 100 projection the spring was measured with 11 different height so that the data included one full circle of spring. See Figure \ref{fig:springarrows}. All those 11 different 2D slices are shown in Figure \ref{fig:slices}. In addition, periodic data sets could be produced by stacking the corresponding {\tt sinogram} and the measurement matrix accordingly.

\clearpage
\section{Contents of the\\ data set}\label{sec:datasets}

The data set contains the following 
data folders/files:\\
{\tt Data\_64x10},\\
{\tt Data\_64x25},\\  
{\tt Data\_256x25},\\
{\tt Data\_64x100},\\
{\tt Data\_256x100},\\
{\tt FilteredBackProjection100.png},\\
{\tt Reconstruction.mp4}\\

\begin{figure*}[ht]
\begin{picture}(300,400)
\put(-215,-50){\includegraphics[width=25cm]{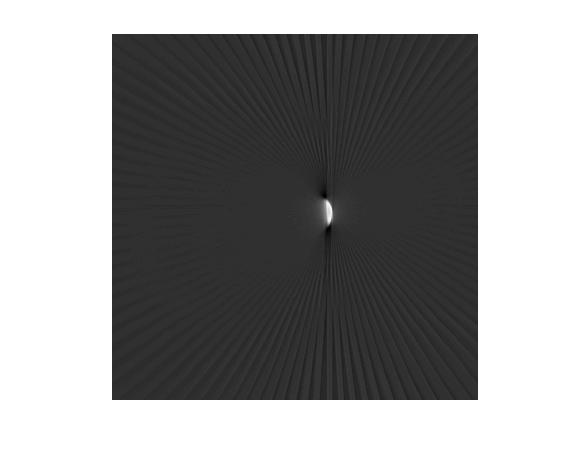}}
\centering
\caption{The filtered back-projection reconstruction ({\tt FilteredBackProjection100.png}) of the spring, with 100 projections.}
\label{fig:fbp}
\end{picture}
\end{figure*}

\noindent
First five data folders include CT sinograms. Folders contain also the corresponding measurement matrices either with the resolution $64 \times 64$ or $256 \times 256$ as spatial resolution and $33$ or $11$ as a temporal resolution in 3D. Those 33 or 11 times instances are obtained by lifting the object little by little at a time between the measurements. The projection angles were same in every time step. Details of these data folders is shown in the Table \ref{tab:datasets}. {\tt FilteredBackProjection100.png} file is the filtered back projection image shown in Figure \ref{fig:fbp} and {\tt Reconstruction.mp4} file is the video made from reconstructions.

\vfill\null
%\columnbreak

\begin{table}[h]
\centering
\begin{tabular}{ l|c|l|c|c|c } 
Folder
&\begin{tabular}{@{}c@{}}Size of \\Matrix {\tt A}\\\end{tabular}
&\begin{tabular}{@{}c@{}}Size of \\{\tt sinogram}\\\end{tabular}
&\begin{tabular}{@{}c@{}}Size of \\{\tt finalSino}\\\end{tabular}
&\begin{tabular}{@{}c@{}}Spatial \\resolution\\\end{tabular}
&\begin{tabular}{@{}c@{}}Temporal \\resolution\\\end{tabular}
\\
\hline
{\tt Data\_64x10} & 970 x 4096 & 1240 x 10 & 97 x 10 & 64 & 33 \\
{\tt Data\_64x25} & 2425 x 4096 & 1240 x 25 & 97 x 25 & 64 & 11 \\
{\tt Data\_256x25} & 9225 x 65536 & 1240 x 25 & 369 x 25 & 256 & 11 \\ 
{\tt Data\_64x100} & 9700 x 4096 & 1240 x 100 & 97 x 100 & 64 & 11 \\
{\tt Data\_256x100} & 36900 x 65536 & 1240 x 100 & 369 x 100 & 256 & 11 \\ 
\hline
\end{tabular}
\caption{Summary of the data folders.}
\label{tab:datasets}
\end{table}
\vfill\eject

\bigskip\noindent
Each folder contains the following variables:
\begin{enumerate}
\item Matrix {\tt A}, the measurement matrix.

\item N {\tt measurement.mat} files that contain the following variables for each N measurements (N equals temporal resolution): 
\begin{enumerate}
\item Matrix {\tt sinogram}, the original measured sinogram.
\item Matrix {\tt finalSino}, the sinogram obtained from {\tt sinogram}.
\end{enumerate}
\end{enumerate}

\bigskip\noindent
More details on the X-ray measurements are described in the Section \ref{sec:Measurements} below.
The model for the CT problem is
\begin{equation}\label{eqn:Axm}
 {\tt A*x=finalSino(:)},
\end{equation}
where {\tt finalSino(:)} denotes the standard vector form of matrix {\tt finalSino} in MATLAB and {\tt x} is the reconstruction in vector form.
In the reconstruction step, the main task is to find that vector {\tt x} that executes the Equation \eqref{eqn:Axm} and possibly meets also some additional regularization requirements. 

\clearpage
\section{X-ray measurements}\label{sec:Measurements}
\begin{figure}[b]
\includegraphics[width=5cm]{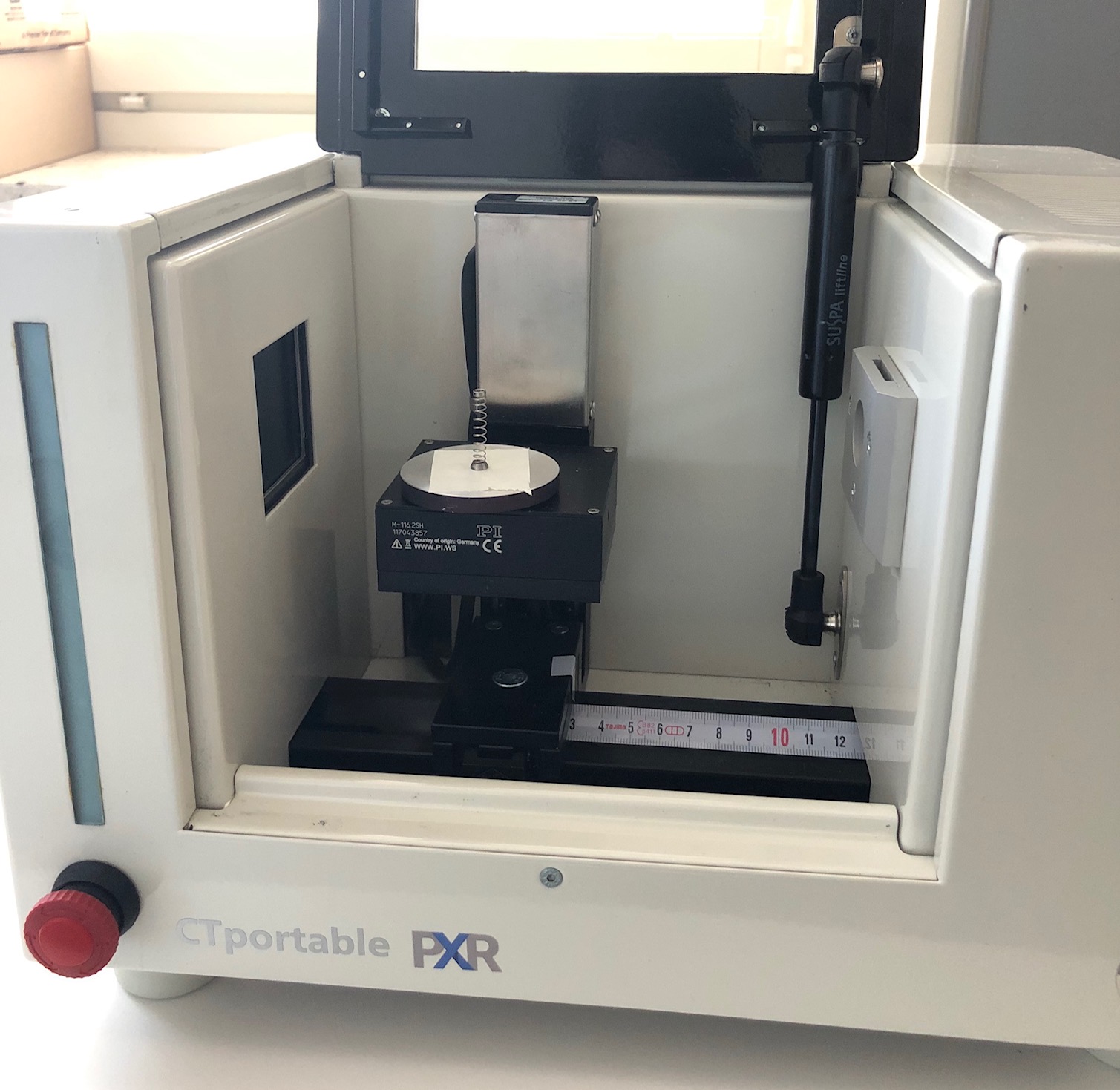}
\centering
\caption{The Procon X-ray CTportable measurement device at Aalto University.}
\label{fig:CTmachine}
\end{figure}

The data in the sinograms are X-ray tomographic (CT) data of a 2D cross-section of the pen-spring measured with Procon X-ray CTportable device shown in 
Figure \ref{fig:CTmachine}.

\begin{figure}[h]
\includegraphics[width=7cm]{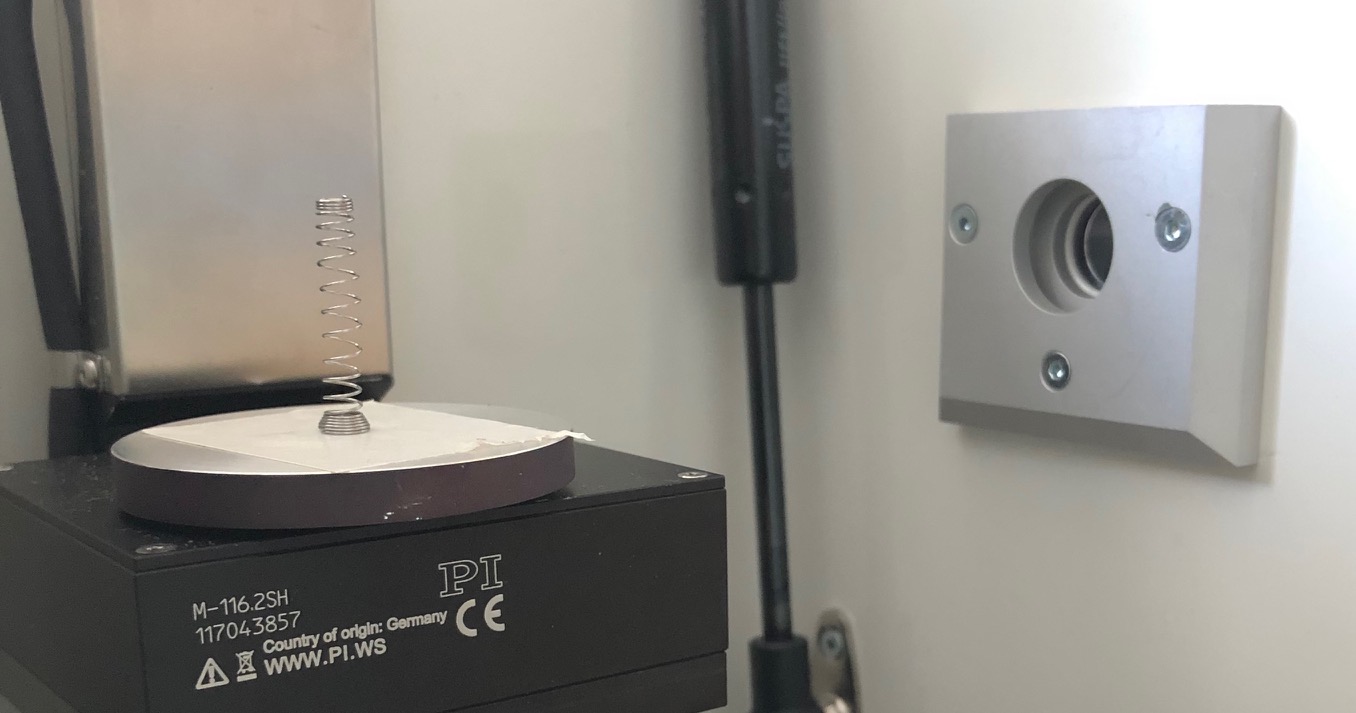}
\centering
\caption{The Procon X-ray CTportable measurement device source.}
\label{fig:source}
\end{figure}

\begin{figure}[h]
\includegraphics[width=7cm]{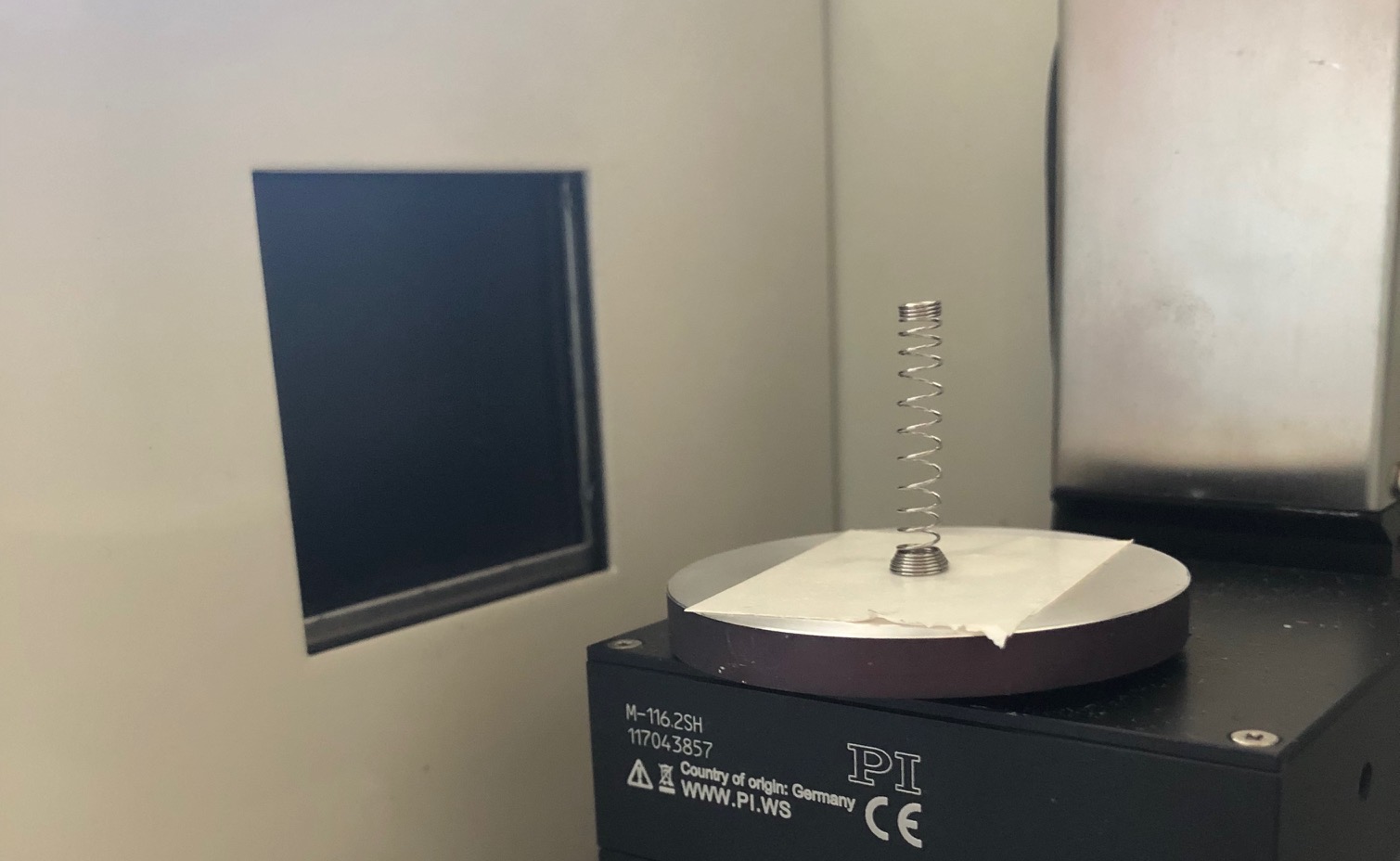}
\centering
\caption{The Procon X-ray CTportable measurement device detector.}
\label{fig:detector}
\end{figure}

\begin{itemize}
\item The X-ray source has 50 kV voltage and 50 W maximum ouput. The source is shown closely in Figure \ref{fig:source}. 
\item The detector has 1 MP pixels and the pixelsize is 48 $\mu$m. The image files generated by the camera were $1024 \times 967$ pixels in size. The detector is shown in Figure \ref{fig:detector}. 
\end{itemize}

\begin{figure}[h]
\includegraphics[width=6cm]{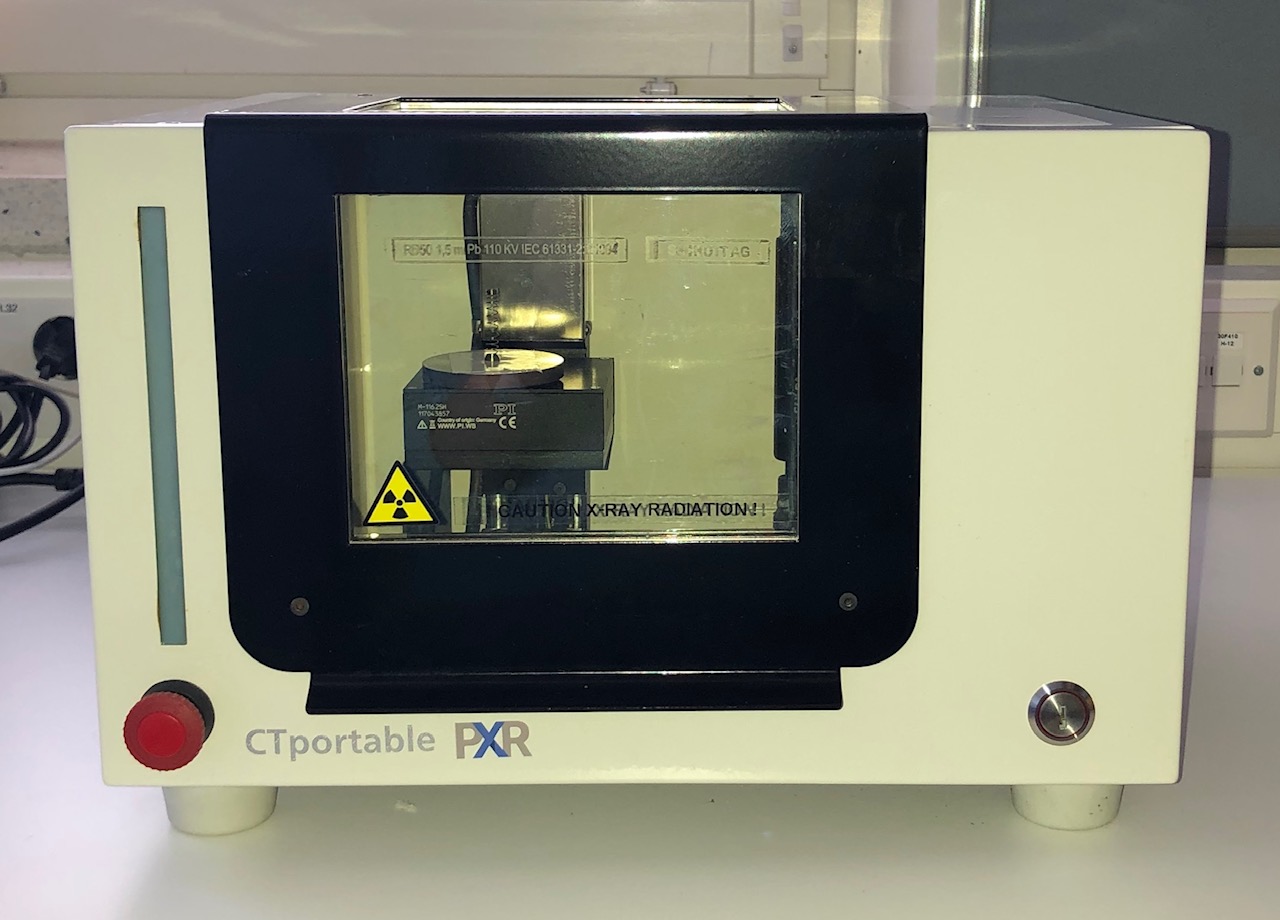}
\centering
\end{figure}

% \vfill\eject 

\begin{figure}[h]
\includegraphics[width=6cm]{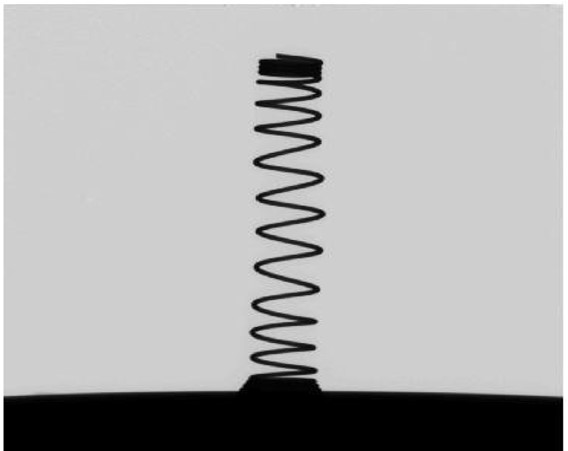}
\centering
\caption{Resulting projection image of the spring.}
\label{fig:spring}
\end{figure}

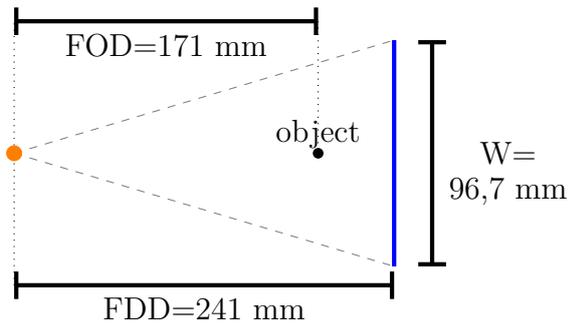
\begin{figure}[h]
\begin{tikzpicture}[scale=1.0]
\draw [ultra thick,blue] (2.5,-1.5) -- (2.5,1.5);
\draw[|-|, ultra thick] (-2.5,1.75) -- (1.5,1.75) node[below,midway]{FOD=171 mm};
\draw [|-|, ultra thick] (-2.5,-1.75) -- (2.5,-1.75)  node[below,midway]{FDD=241 mm}; 
\draw[|-|, ultra thick] (3.0,-1.5) -- (3.0,1.5); % vertical
\draw[dotted] (-2.5,-1.75) -- (-2.5,1.75); % vertical
\draw[dotted] (1.5,1.75) -- (1.5,0.0); % vertical
\draw[dashed,gray] (-2.5,0.0) -- (2.5,-1.5);
\draw[dashed,gray] (-2.5,0.0) -- (2.5,1.5); 
\draw (4,0.0) node{W=}; 
\draw (4,-0.5) node{96,7 mm};
\fill[thick] (1.5,0.0) circle (2pt);
\draw (1.5,0.25) node{object};
\fill[thick, orange] (-2.5,0.0) circle (3pt);
\end{tikzpicture}
\bigskip
\caption{Geometry of the measurement setup. Here FOD and FDD denote the focus-to-object distance and the focus-to-detector distance, respectively; the black dot object is the center-of-rotation. The width of the detector (the blue line) is denoted by W. The orange dot is the X-ray source. To increase clarity, the $x$-axis and $y$-axis in this image are not in scale.}\label{geometry}
\end{figure}

\clearpage\noindent
The measurement geometry is shown in Figure \ref{geometry}. A set of 360 fan-beam projections with resolution $1024 \times 967$ 
was measured. The exposure time was 400 ms, X-ray tube acceleration voltage 50 kV and tube current 400 mA. See Figure~\ref{fig:spring} for an example of the resulting projection images. 

\begin{figure}[h]
\begin{picture}(100,270)
\put(50,320){\color{gray}\line(1,-3){100}}
\put(50,320){\color{gray}\line(-1,-3){100}}

\put(50,315){\color{orange}\circle*{20}}
\put(0,30){\line(1,0){100}}
\put(0,55){\line(1,0){100}}
\put(0,80){\line(1,0){100}}
\put(0,105){\line(1,0){100}}
\put(0,130){\line(1,0){100}}
\put(0,30){\line(0,1){100}}
\put(25,30){\line(0,1){100}}
\put(50,30){\line(0,1){100}}
\put(75,30){\line(0,1){100}}
\put(100,30){\line(0,1){100}}
\put(8,115){$x_1$}
\put(8,90){$x_2$}
\put(10,62){$\vdots$}
\put(6,40){$x_N$}
\put(26,115){$x_{\scriptscriptstyle N+1}$}
\put(26,90){$x_{\scriptscriptstyle N+2}$}
\put(35,62){$\vdots$}
\put(28,40){$x_{2N}$}
\put(55,115){$\cdots$}
\put(55,90){$\cdots$}
\put(55,62){$\ddots$}
\put(55,40){$\cdots$}
\put(78,40){$x_{N^2}$}

\put(-50,0){\color{blue}\line(1,0){200}}
\put(-50,25){\color{blue}\line(1,0){200}}
\put(-50,0){\color{blue}\line(0,1){25}}
\put(-25,0){\color{blue}\line(0,1){25}}
\put(0,0){\color{blue}\line(0,1){25}}
\put(25,0){\color{blue}\line(0,1){25}}
\put(100,0){\color{blue}\line(0,1){25}}
\put(125,0){\color{blue}\line(0,1){25}}
\put(150,0){\color{blue}\line(0,1){25}}
\put(-45,10){$m_1$}
\put(-20,10){$m_2$}
\put(5,10){$\cdots$}
\put(30,10){$\cdots$}
\put(83,10){$\cdots$}
\put(105,10){$\cdots$}
\put(130,10){$m_N$}
\end{picture}
\caption{The organization of the pixels in the sinograms {\tt m}\,=\,$[m_1,m_2,\ldots,m_{60K}]^T$ and reconstructions {\tt x}\,=\,$[x_1,x_2,\ldots,x_{N^2}]^T$with $N=64$ or $N=256$ . The picture shows the organization for the first projection; after that in the full angular view case, the target takes steps ($3.6$ degree step for 10 projection data and $36$ degree step for 100 projection data) counter-clockwise (or equivalently the source and detector take steps clockwise) and the following columns of {\tt m} are determined in an analogous manner.}\label{fig:pixelDemo}
\end{figure}
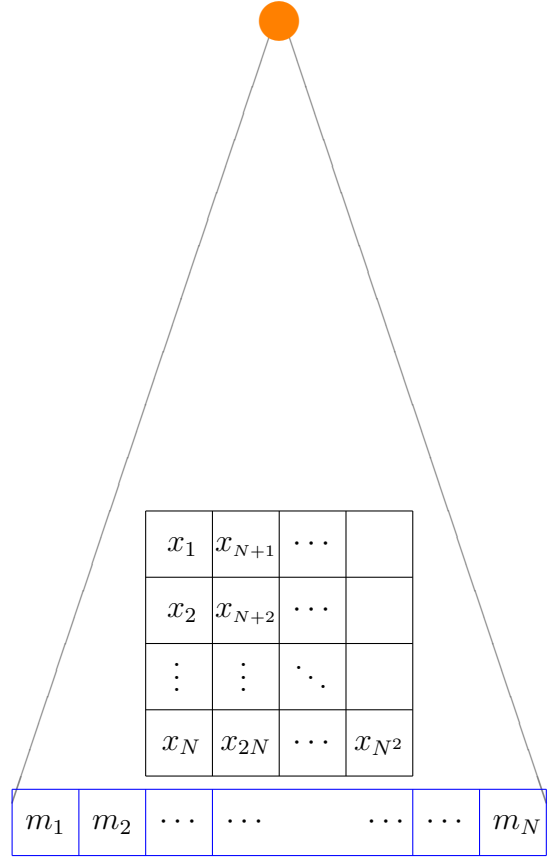

From the 2D projection images, the middle rows (row 483) corresponding to the central horizontal cross-section of the spring target were taken to form a
fan-beam sinogram of resolution $1024 \times 10$ or $100$. These sinograms were further down-sampled by binning, taken logarithms and normalized to obtain the {\tt finalSino} in all the files specified in Section~\ref{sec:datasets}. The organization of the pixels in the sinograms and the reconstructions is illustrated in Figure~\ref{fig:pixelDemo}.

\clearpage

\clearpage
\section{3D reconstruction}\label{sec:video}
The video of the reconstruction of the target is available in the data set in Zenodo.

% \bigskip
% \includemedia[width=1\linewidth,height=1\linewidth,activate=pageopen,
% passcontext,
% transparent,
% addresource=videoslow.mp4,
% flashvars={source = videoslow.mp4}
% ]{\includegraphics[width=2\linewidth]{frame5}}{VPlayer.swf}\label{video}

\section*{Acknowledgement}
This work was funded by the Academy of Finland and by the School of Electrical Engineering, Aalto University, Finland.

\bibliographystyle{ieeetr}
\bibliography{ref}

\end{document}